\documentclass[twocolumn,showpacs,superscriptaddress,amsmath,amssymb,aps,pra]{revtex4-1}
\usepackage{graphicx}
\usepackage{CJKutf8}
\usepackage{dcolumn}
\usepackage{amsmath,bm}
\usepackage{epstopdf}
\usepackage[mathlines]{lineno}
\usepackage{color}
\usepackage[colorlinks=true,linkcolor=blue,citecolor=magenta,urlcolor=cyan]{hyperref}

\begin{document}
\title{Spin manipulation and spin dephasing in quantum dot integrated with a slanting magnetic field}
\author{Rui\! Li~(\begin{CJK}{UTF8}{gbsn}李睿\end{CJK})}
\email{ruili@ysu.edu.cn}
\affiliation{Key Laboratory for Microstructural Material Physics of Hebei Province, School of Science, Yanshan University, Qinhuangdao 066004, China}

\begin{abstract}
A slanting magnetic field is usually used to realize a slight hybridization between the spin and orbital degrees of freedom in a semiconductor quantum dot, such that the spin is manipulable by an external oscillating electric field. Here we show that, the longitudinal slanting field mediates a longitudinal driving term in the electric-dipole spin resonance, such that the spin population inversion exhibits a modulated Rabi oscillation. Fortunately, we can reduce this modulation by increasing the static magnetic field. The longitudinal slanting field also mediates a spin-1/f-charge noise interaction, which causes the pure dephasing of the spin qubit. Choosing proper spectrum function strength, we find the spin dephasing time is about $T^{*}_{2}=20$ $\mu$s and the spin echo time is about $T^{\rm echo}_{2}=100$ $\mu$s in a Si quantum dot. We also propose several strategies to alleviate the spin dephasing, such as lowering the experimental temperature, reducing the quantum dot size, engineering the slanting field, or using the dynamical decoupling scheme.
\end{abstract}
\date{\today}
\maketitle

\section{Introduction}
Electron spin confined in a semiconductor quantum dot is a promising qubit candidate because of both the long dephasing time and the relative convenience for scalability~\cite{Loss1998,Hanson2007}. The spin dephasing time can be as long as a millisecond in isotopically purified Si quantum dot~\cite{veldhorst2014,veldhorst2015}. While in III-V semiconductor quantum dots, such as GaAs, the spin dephasing time is in the microsecond region~\cite{petta2005}, limited mainly by the hyperfine interaction between the electron and lattice nuclear spins~\cite{Yao2006,Cywinski2009}. Single qubit manipulation in the quantum dot can be achieved via either the electron spin resonance~\cite{koppens2006} or the electric-dipole spin resonance (EDSR)~\cite{Rashba2003,Golovach2006,LiRui2013,nowack2007,Nadj2012}. Two qubit manipulation can be naturally achieved by using the exchange interaction in a double quantum dot~\cite{Burkard1999,Hu2000}.

The manipulation time $T_{\rm Rabi}$ and the dephasing time $T^{*}_{2}$ are two important time scales for the qubit~\cite{Buluta2011}. The values of these two quantities determine whether a qubit candidate is suitable for quantum computing. An ideal quantum computer requires that enough number (about 1000) of single qubit manipulations should be completed in the qubit dephasing time~\cite{ladd2010}. Dephasing is a leading obstacle limiting all potential applications of the qubit. In order to alleviate the qubit suffering from dephasing caused by environmental noises, we should first understand various possible dephasing mechanisms~\cite{chan2018}.

There are both no internal spin-orbit coupling and negligible lattice nuclear spins in isotopically purified 28Si, such that Si quantum dot is expected to be one of the most feasible platforms for quantum computing~\cite{veldhorst2014,veldhorst2015}. The spin qubit in Si quantum dot is so separate from the external environment that single qubit manipulation becomes relatively inconvenient. Electron spin resonance in a quantum dot is proved to be technically challenging~\cite{koppens2006}. A feasible way is to integrate the quantum dot with a slanting magnetic field~\cite{pioro2008,Brunner2011,kawakami2014,Chesi2014,Forster2015,Scarlino2015}, such that single spin manipulation can be achieved via EDSR. However, as observed in experiments, the slanting field also brought the 1/f charge noise to the spin qubit~\cite{Kawakami2016,yoneda2018}. 1/f charge noise commonly exists in many nano-structures~\cite{Dutta1981,Weissman1988,Paladino2014}, and it has also been regarded as the main noise source that causes the dephasing of the qubit, such as Josephson qubit~\cite{Astafiev2004,You2007,bylander2011}, quantum dot charge qubit~\cite{Petersson2010,Shi2013}, spin qubit~\cite{chan2018,Kha2015}, singlet-triplet qubit~\cite{Culcer2009,Hu2006,Gamble2012}, etc.

In this paper, we study the slanting field mediated spin manipulation and spin dephasing in a Si quantum dot. In the spin manipulation via EDSR, the transverse slanting field mediates a transverse driving term which contributes to the periodic oscillation of the spin population inversion, while the longitudinal slanting field mediates a longitudinal driving term which gives a modulation to the spin population inversion. Fortunately, the effect of the modulation can be reduced by applying a large Zeeman field to the quantum dot. The pure dephasing is caused by the longitudinal spin-1/f-charge noise interaction, which is also mediated by the longitudinal slanting field. We propose prolonging the spin dephasing time by reducing the quantum dot size, lowering the experimental temperature, reducing the longitudinal slanting field, or using a dynamical decoupling scheme~\cite{Uhrig2007}. Under eight pulse sequences, the spin dephasing time $T_{2}$ can be prolonged to the sub-millisecond region. Finally, because the upper bound of the 1/f charge noise spectrum is usually less than the qubit level spacing in the quantum dot, the 1/f charge noise cannot contribute to the spin relaxation.

\section{The model}
\begin{table}
\centering
\caption{\label{tab}The parameters of the Si quantum dot used in our calculations. The values are taken from Ref.~\onlinecite{yoneda2018}}
\begin{ruledtabular}
\begin{tabular}{ccccccc}
$m/m_{0}$\footnote{$m_{0}$ is the free electron mass}&$g$&$B_{0}$~(T)&$\omega_{0}$~(THz)\footnote{$r_{0}=\sqrt{\hbar/(m\omega_{0})}=20$ nm}&$b_{t}$\footnote{in unit of (mT/nm), $z_{0}=g\mu_{B}b_{t}/(2m\omega^{2}_{0})=2.431\times10^{-2}$nm}&$b_{l}$\footnote{in unit of (mT/nm), $y_{0}=g\mu_{B}b_{l}/(2m\omega^{2}_{0})=0.4862\times10^{-2}$nm}&{\it T} (mK)\\
$0.2$&$2$&$0.5$&1.447&$1.0$&$0.2$&100
\end{tabular}
\end{ruledtabular}
\end{table}

We consider a realistic quantum dot model which is intimately related to the experimental situations demonstrated recently~\cite{yoneda2018,Yoneda2014}. The quantum dot has a two-dimensional harmonic confining potential on the $yz$ plane and is exposed to both static and slanting magnetic fields. The slanting field, which is used to assist the spin manipulation via an external electric-field, is created by covering a Co micromagnet on the quantum dot~\cite{yoneda2018,Yoneda2014,Neumann2015,Wu2014}. The model under consideration reads
\begin{equation}
H=\frac{p^{2}_{y}+p^{2}_{z}}{2m}+\frac{1}{2}m\omega^{2}_{0}(y^{2}+z^{2})+\frac{g\mu_{B}({\bf B}_{0}+{\bf B}_{\rm m})\cdot\boldsymbol{\sigma}}{2},
\end{equation}
where $m$ is the effective electron mass, $\omega_{0}$ is the frequency of the harmonic confining potential [the quantum dot characteristic length $r_{0}=\sqrt{\hbar/(m\omega_{0})}$], ${\bf B}_{0}=(0,0,B_{0})$ is an in-plane static field applied long the $z$ direction, and ${\bf B}_{\rm m}=(B^{x}_{\rm m},B^{y}_{\rm m},B^{z}_{\rm m})$ is the stray field induced by the Co micromagnet. One can expand the stray field up to the linear terms using Taylor's formula
\begin{eqnarray}
B^{x}_{\rm m}(z)&=&B^{x}_{\rm m}(0)+b_{t}z,\nonumber\\
B^{y}_{\rm m}(z)&=&B^{y}_{\rm m}(0)+b_{l}z,\nonumber\\
B^{z}_{\rm m}(x,y)&=&B^{z}_{\rm m}(0)+b_{l}y+b_{t}x,\label{eq_slantingfield}
\end{eqnarray}
where $b_{t}$ and $b_{l}$ are the slopes of the transverse and longitudinal fields~\cite{yoneda2018,Yoneda2014}, respectively. One can check that the above stray field does not violate Maxwell's equations $\nabla\cdot{\bf B}_{\rm m}=0$ and $\nabla\times{\bf B}_{\rm m}=0$ for a static system. The small $x,y$-components $B^{x,y}_{m}(0)$ of the stray field are neglected from consideration and the $z$-component $B^{z}_{m}(0)$ can be absorbed to the static Zeeman field $B_{0}$. After the above linear approximation, the quantum dot Hamiltonian can be written as
\begin{eqnarray}
H&=&\frac{p^{2}_{y}+p^{2}_{z}}{2m}+\frac{m\omega^{2}_{0}}{2}(y^{2}+z^{2}+2y_{0}y\sigma^{z})+\Delta\sigma^{z}\nonumber\\
&&+m\omega^{2}_{0}z\sqrt{z^{2}_{0}+y^{2}_{0}}(\sigma^{x}\cos\theta_{0}+\sigma^{y}\sin\theta_{0}),\label{eq_model2}
\end{eqnarray}
where $y_{0}=g\mu_{B}b_{l}/(2m\omega^{2}_{0})$ and $z_{0}=g\mu_{B}b_{t}/(2m\omega^{2}_{0})$ characterize the length scale of the  longitudinal and transverse gradient fields, respectively, $\theta_{0}=\arctan(y_{0}/z_{0})$ and $\Delta=g\mu_{B}B_{0}/2$ is half of the Zeeman splitting. It should be noted that the vector potential $\textbf{A}=\textbf{B}_{0}\times\textbf{r}/2$ is perpendicular to the $yz$ plane, such that there are no vector potential components in the Hamiltonian (\ref{eq_model2}), i.e., $p_{y}=-i\hbar\partial_{y}$ and $p_{z}=-i\hbar\partial_{z}$. Also, we have assumed the quantum dot lies on the $x=0$ plane.

In line with the experimental investigation~\cite{yoneda2018}, here we choose Si as our quantum dot material. In our following calculations, unless otherwise stated, the parameters chosen are listed in table~\ref{tab}.

\section{Slanting field mediated electric-dipole spin resonance}

\begin{figure}
\centering
\includegraphics[width=8.5cm]{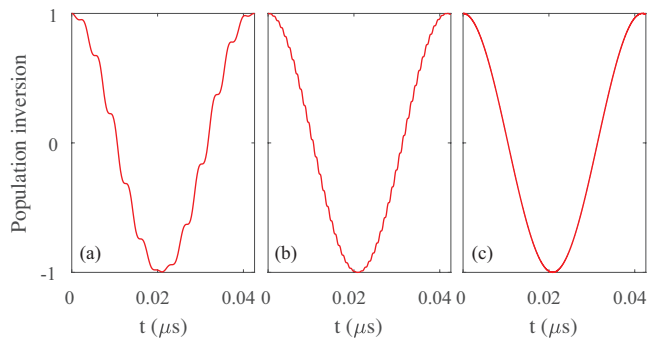}
\caption{\label{fig_Rabi_Si}EDSR described by the driving Hamiltonian (\ref{eq_edsr}) under the resonant condition $\hbar\omega=2\Delta$. The qubit population inversion is defined as $|c_{\Uparrow}(t)|^{2}-|c_{\Downarrow}(t)|^{2}$ for state: $|\varphi(t)\rangle=c_{\Uparrow}(t)|\!\!\Uparrow\rangle+c_{\Downarrow}(t)|\!\!\Downarrow\rangle$.  The qubit is initially in state $|\varphi(0)\rangle=|\!\!\Uparrow\rangle$ and the driving strength is chosen as $E_{y}=E_{z}=4000$ V/m. The results at the external fields of $B_{0}=0.005$ T (a), $B_{0}=0.02$ T (b), and $B_{0}=0.5$ T (c).}
\end{figure}

The manipulation of the quantum-dot spin qubit is usually achieved via EDSR. Quantum-dot EDSR can be mediated by internal spin-orbit coupling~\cite{Rashba2003,Golovach2006,LiRui2013,nowack2007,Nadj2012,Khomitsky2012,Nowak2013,Romhanyi2015}, electron-nuclear hyperfine interaction~\cite{Laird2007,Rashba2008,LiRui2016}, and external slanting magnetic field~\cite{Tokura2006,Rancic2016}. In the earlier seminal work of Tokura and co-workers~\cite{Tokura2006}, only a transverse slanting field is proposed to mediate the EDSR. However, under realistic experimental circumstance, the micromagnet brings no only the transverse but also the longitudinal slanting fields to the quantum dot~\cite{pioro2008,Brunner2011,yoneda2018,Yoneda2014} [see Eq.~(\ref{eq_model2})]. Here we examine the impacts of the longitudinal slanting field on the spin manipulation.

Under the external electric-field driving, an additional electric-dipole interaction term $e\textbf{E}\cdot\textbf{r}\cos\omega\,t$ should be added to Hamiltonian (\ref{eq_model2}). When we focus only on the qubit Hilbert space spanned by $|\!\!\Uparrow\rangle\equiv|\Psi_{0,0,\uparrow}\rangle$ and $|\!\!\Downarrow\rangle\equiv|\Psi_{0,0,\downarrow}\rangle$, the electric-driving Hamiltonian can be reduced to the form of a two-level atom interacting with a classical field~\cite{scully1999quantum} (for details see Appendix \ref{appendix_A})
\begin{eqnarray}
H_{\rm dr}&=&\Delta\tau^{z}-eE_{y}y_{0}\tau^{z}\cos\omega\,t\nonumber\\
&&-eE_{z}\sqrt{z^{2}_{0}+y^{2}_{0}}(\tau^{x}\cos\theta_{0}+\tau^{y}\sin\theta_{0})\cos\omega\,t,\label{eq_edsr}
\end{eqnarray}
where $\tau^{z}=|\!\!\Uparrow\rangle\langle\Uparrow\!\!|-|\!\!\Downarrow\rangle\langle\Downarrow\!\!|$, $\tau^{x}=|\!\!\Uparrow\rangle\langle\Downarrow\!\!|+|\!\!\Downarrow\rangle\langle\Uparrow\!\!|$, $\tau^{y}=-i|\!\!\Uparrow\rangle\langle\Downarrow\!\!|+i|\!\!\Downarrow\rangle\langle\Uparrow\!\!|$, $E_{y}$ and $E_{z}$ are the $y$ and $z$ components of the driving-field, respectively, and $\omega$ is the frequency of the driving-field. This Hamiltonian is slightly different from the standard Rabi oscillation Hamiltonian in quantum optics~\cite{scully1999quantum} because of the presence of the second term, which is induced by the longitudinal slanting field given in Eq.~(\ref{eq_model2}).

Let us examine the influence of the longitudinal driving term [the second term in Eq.~(\ref{eq_edsr})] on the spin manipulation. Similar to the standard Rabi oscillation, the qubit is initially prepared in state $|\varphi(0)\rangle=|\!\!\Uparrow\rangle$. When the frequency of the driving field matches the qubit level spacing $\hbar\omega=2\Delta$, the spin population inversion is obtained by numerically solving the time-dependent Schr\"odinger equation governed by Hamiltonian (\ref{eq_edsr}). We find that, at small external magnetic field such as $B_{0}=0.005$ T, there is an apparent modulation on the spin population inversion [see Fig.~\ref{fig_Rabi_Si}(a)]. When the magnetic field is increased to $B_{0}=0.02$ T, the modulation becomes relative small[see Fig.~\ref{fig_Rabi_Si}(b)]. When the external magnetic field is large enough, such as $B_{0}=0.5$ T, the modulation becomes negligible (almost invisible) [see Fig.~\ref{fig_Rabi_Si}(c)]. Anyway, one can reduce the modulation via increasing the external magnetic field $B_{0}$. This is very reasonable, the longitudinal driving term can be regarded as a time-dependent Zeeman field applied to the spin qubit $(\Delta-eE_{y}y_{0}\cos\omega\,t)\tau^{z}$. The larger the static magnetic field, the smaller the relative ratio $eE_{y}y_{0}/\Delta$, hence the smaller the effect of the longitudinal driving term.

Next, let us analyze the strength of the Rabi frequency, which characterizes the qubit manipulation time. Note that the qubit is encoded to the lowest two energy levels of the quantum dot. Although the qubit Hilbert space is well separated from the other higher orbital levels in the quantum dot, i.e., the Zeeman splitting $2\Delta_{B_{0}=0.5~{\rm T}}$ (0.058 meV) is much smaller than the orbital splitting $\hbar\omega_{0}$ (0.95 meV), there still exist leakages from the qubit Hilbert space to the higher orbital states under the strong field driving. The spin dynamics in this case are totally nontrivial, and one has to consider the multi-level effects in the EDSR~\cite{Khomitsky2012}. In order to avoid the electron being excited to higher orbital states, here the electric field strength is constrained to $|\textbf{E}|\ll(\hbar\omega_{0})/(er_{0})=4.769\times10^{4}$ V/m. This result gives an upper bound on the Rabi frequency in our model $\Omega_{R}\ll\,eE_{\rm max}\sqrt{z^{2}_{0}+y^{2}_{0}}/h=286$ MHz, and agrees qualitatively well with the experimental observations~\cite{yoneda2018,Takedae2016}.

\section{\label{sec_IV}Charge noise induced pure-dephasing}

1/f charge noise has been observed in many quantum nano-structures~\cite{Dutta1981,Weissman1988,Paladino2014}, and it has also been regarded as the main noise limiting the dephasing time of many qubit candidates~\cite{Astafiev2004,You2007,bylander2011,Petersson2010,Shi2013,Kha2015,Culcer2009,Hu2006,Gamble2012,lirui2018a}. The physical origin of the charge fluctuation spectrum with 1/f distribution is still unclear, and many theoretical models have been proposed~\cite{Paladino2014}. Here we just assume that the charge field has a spectrum function $A^{2}/\omega$, and the value of $A$ is chosen to fit well with the experimental observation.

We assume the fluctuating charge field has a similar form as that of the vacuum electromagnetic field~\cite{scully1999quantum}
\begin{equation}
\textbf{E}(\textbf{r})=\sum_{k}\Xi_{k}\vec{e}_{k}(a_{k}e^{i\vec{k}\cdot\vec{r}}+a^{\dagger}_{k}e^{-i\vec{k}\cdot\vec{r}}),\label{eq_chargefield}
\end{equation}
where $\Xi_{k}$ is the charge field in the wavevector space, $\vec{e}_{k}$ is a unit vector, and $\vec{k}$ is the wavevector. The transverse character of the electromagnetic field gives rise to $\vec{e}_{k}\cdot\vec{k}=0$~\cite{scully1999quantum}. In order to simplify the complexity of the problem, we further assume the wave is propagating along the $x$ direction: $\vec{k}=k\vec{e}_{x}\perp\,yz$ plane, such that $\vec{e}_{k}$ is an in-plane unit vector, hence $\textbf{E}(\textbf{r})=\sum_{k}\Xi_{k}\vec{e}_{k}(a_{k}+a^{\dagger}_{k})$ (the quantum dot is confined on the $x=0$ plane). Replacing the classical field in Eq.~(\ref{eq_edsr}) with the above quantized electric-field, we obtain the total Hamiltonian describing the interaction between the spin qubit and the charge noise
\begin{eqnarray}
H_{\rm tot}&=&\Delta\tau^{z}-\sum_{k}e\Xi_{k}y_{0}\tau^{z}(a_{k}+a^{\dagger}_{k})\cos\Theta+\sum_{k}\hbar\omega_{k}a^{\dagger}_{k}a_{k}\nonumber\\
&&-\sum_{k}e\Xi_{k}(z_{0}\tau^{x}+y_{0}\tau^{y})(a_{k}+a^{\dagger}_{k})\sin\Theta,\label{eq_decoherence}
\end{eqnarray}
where $\Theta$ is the azimuth of the charge field on the $yz$ plane. The exact value of $\Theta$ is unknown, such that it is reasonable to average over all possible angle $\Theta$ for the obtained physical quantities, e.g., $\Gamma(t)\equiv\langle\Gamma(t)\rangle_{\Theta}=\int^{2\pi}_{0}\Gamma(t)d\Theta/2\pi$.

\begin{figure}
\centering
\includegraphics{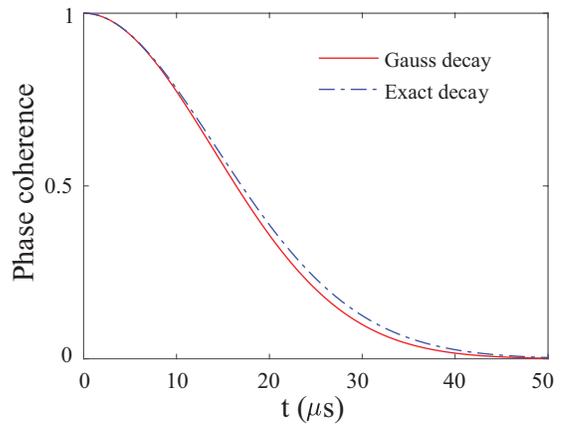}
\caption{\label{fig_dephasing}The pure dephasing of the spin qubit due to the $1/f$ charge noise. We have chosen the noise spectrum strength $A_{r_{0}=20~{\rm nm},T=100~{\rm mK}}=35$ MHz in order to fit well with the experimental observation~\cite{yoneda2018}.}
\end{figure}

The pure-depasing of the qubit is caused by the longitudinal coupling between the qubit and the charge noise as illustrated by the second term in Eq.~(\ref{eq_decoherence}). This term can been traced back to the longitudinal slanting term in Eq.~(\ref{eq_model2}). If we model the qubit dephasing as ${\rm exp}\left[-\Gamma_{\rm ph}(t)\right]$, the decaying factor be written as~\cite{Palma1996}
\begin{equation}
\Gamma_{\rm ph}(t)=2\frac{y^{2}_{0}}{r^{2}_{0}}\int^{\omega_{\rm max}}_{\omega_{\rm min}}d\omega\,S(\omega)\frac{\sin^{2}(\omega\,t/2)}{(\omega/2)^{2}},\label{eq_dephasingrate}
\end{equation}
where $\omega_{\rm min (max)}$ is the lower (upper) bound of the noise frequency, and the spectrum function is defined as
\begin{eqnarray}
S(\omega)&=&\sum_{k}\frac{e^{2}r^{2}_{0}\Xi^{2}(\omega)[2n(\omega)+1]}{2\hbar^{2}}\delta(\omega-\omega_{k})\nonumber\\
&\approx&\sum_{k}\frac{e^{2}r^{2}_{0}\Xi^{2}(\omega)k_{B}T}{\hbar^{3}\omega}\delta(\omega-\omega_{k})\equiv\frac{A^{2}_{r_{0},T}}{\omega},\label{eq_noisespectrum}
\end{eqnarray}
with $A_{r_{0},T}$ being a parameter characterizing the strength of the charge noise. The lower bound of the noise spectrum is about $\omega_{\rm min}\approx\,10^{-2}$ Hz~\cite{yoneda2018}, and the upper bound of the noise spectrum is about $\omega_{\rm max}\approx\,5\times10^{5}$ Hz~\cite{yoneda2018}. We have also included the temperature effect in deriving Eq.~(\ref{eq_noisespectrum}) by writing the Bose occupation number as $n(\omega)=1/\left[{\rm exp}(\hbar\omega/k_{B}T)-1\right]\approx\,k_{B}T/(\hbar\omega)\gg1$, under the realistic temperature~\cite{yoneda2018} ($T=100$ mK) for all the low frequency noise modes ($\omega_{\rm max}\sim0.004$ mK). Note that $A_{r_{0},T}$ has the dimension of the frequency, in order to fit well with the experimental observed dephasing time $T^{*}_{2}\approx20\,\mu$s~\cite{yoneda2018}, we have chosen $A_{r_{0}=20~{\rm nm},T=100~{\rm mK}}=35$ MHz (see Fig.~\ref{fig_dephasing}). It is instructive to see for the time scale $t<1/\omega_{\rm max}=2 \mu$s, we can write the dephasing factor as the following simple form (a similar version of Ref.~\onlinecite{Schriefl2006})
\begin{equation}
\Gamma_{\rm ph}(t)=2A^{2}_{r_{0},T}t^{2}\frac{y^{2}_{0}}{r^{2}_{0}}\ln\frac{\omega_{\rm max}}{\omega_{\rm min}}.\label{eq_gaussdecay}
\end{equation}
Thus, the qubit dephasing at short time must be a Gauss decay. Actually, for time scale larger than $t>1/\omega_{\rm max}$ in our model, we find that the difference between the Gauss decay (\ref{eq_gaussdecay}) and the exact decay (\ref{eq_dephasingrate}) is very small (see Fig.~\ref{fig_dephasing}).

Let us discuss on the spectrum function defined in Eq.~(\ref{eq_noisespectrum}). Although our derivation of the spectrum function with 1/$\omega$ distribution has been made plausible, the difficulty lies in choosing reasonable $\Xi_{k}$ such that the second expression can be written as the third expression in the last line of Eq.~(\ref{eq_noisespectrum}). Actually, the physical mechanism of the charge spectrum with 1/$\omega$ distribution is still unclear~\cite{Paladino2014}. Here, we give a simple argument to realize the 1/f spectrum function. Note that the wavevector $\vec{k}$ is perpendicular to the $yz$ plane, and for the electromagnetic wave we have the dispersion relation $\omega_{k}=ck$, where $c$ is the speed of light. We make the following replacement in Eq.~(\ref{eq_noisespectrum}) $\sum_{k}\rightarrow\int\,d\omega_{k}L/(\pi\,c)$, where $L$ is the length of the space in the $x$ dimension ($V=L^{3}$). It is suggested that the charge field of wavevector $\Xi_{k}$ should be a constant $\Xi_{k}\equiv\Xi$, which is in stark contrast with that of the vacuum electromagnetic field~\cite{scully1999quantum}. Hence the spectrum function can be written as
\begin{equation}
S(\omega)=\frac{e^{2}r^{2}_{0}\Xi^{2}Lk_{B}T}{\pi\,c\hbar^{3}\omega},
\end{equation}
which is indeed of the 1/$\omega$ form. Note that the linear temperature dependence of the spectrum function is consistent with both theoretical~\cite{Dutta1981,Culcer2009} and experimental~\cite{Jung2004} investigations. Although we only study the low-frequency 1/f charge noise, it is still of interest to discuss the spectrum function in the high-frequency region under this argument. Note that the first line of Eq.~(\ref{eq_noisespectrum}) is valid in all frequency range. For the high-frequency noise modes $\hbar\omega\gg\,k_{B}T$, $n(\omega)=1/\left[{\rm exp}(\hbar\omega/k_{B}T)-1\right]\approx0$. Hence, the spectrum function in the high-frequency region should be
\begin{equation}
S(\omega)=\frac{e^{2}r^{2}_{0}\Xi^{2}L}{2\pi\,c\hbar^{2}}.
\end{equation}
This spectrum function is irrelevant to the frequency $\omega$. In the noise theory, noise with this kind of spectrum is called white noise~\cite{Paladino2014}.

\section{Prolong the dephasing time}
The dephasing time $T^{*}_{2}$ is an important time scale for the qubit~\cite{Buluta2011}. A long dephasing time is always appreciated for almost all qubit candidates. Based on the spin dephasing theory built in the above section, here we study how to prolong the spin dephasing time in a Si quantum dot.

\begin{figure}
\includegraphics{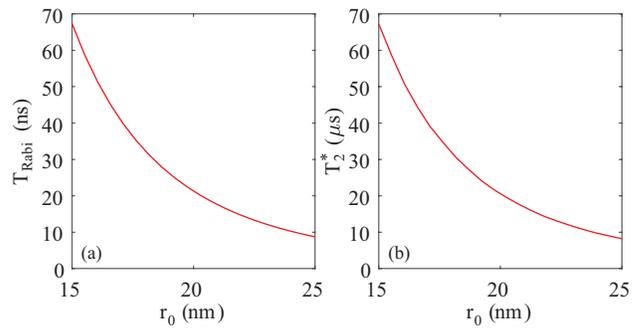}
\caption{\label{fig_dephasingvsr0}The spin manipulation time (a) and the spin dephasing time (b) as a function of the quantum dot characteristic length $r_{0}$. The manipulation time is defined as $T_{\rm Rabi}=h/(2eE_{z}\sqrt{z^{2}_{0}+y^{2}_{0}})$, where $E_{z}=4000$ V/m, and the dephasing time $T^{*}_{2}$ is solved from $\Gamma_{\rm ph}(T^{*}_{2})=1$.}
\end{figure}

The first intuitional approach is to reduce the quantum dot characteristic length $r_{0}$~\cite{Bermeister2014}. The characteristic length is related to the electric dipole moment of the quantum dot, such that reducing $r_{0}$ obviously reduces the effective coupling between the spin and the charge noise in Eq.~(\ref{eq_decoherence}). However, the coupling between the spin and the classical field, i.e., the Rabi frequency in Eq.~(\ref{eq_edsr}), is reduced simultaneously. Therefore, reducing $r_{0}$ not only increases the dephasing time $T^{*}_{2}$ [see Fig.~\ref{fig_dephasingvsr0}(b)] but also increases the Rabi manipulation time $T_{\rm Rabi}$ [see Fig.~\ref{fig_dephasingvsr0}(a)]. The $r_{0}$ dependence of the dephasing can be roughly written as ${\rm T}^{*}_{2}\propto\,r^{-4}_{0}$. From this viewpoint, reducing $r_{0}$ may not be an effective way to prolong the dephasing time. Note that the spin dephasing time $T^{*}_{2}$ is obtained by solving $\Gamma_{\rm ph}( T^{*}_{2})=1$ in Eq.~(\ref{eq_dephasingrate}).

\begin{figure}
\includegraphics{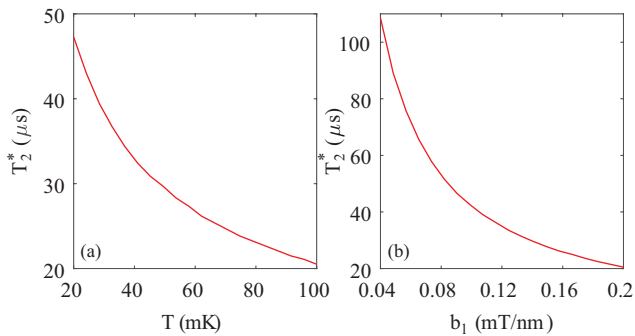}
\caption{\label{fig_dephasingvstemperature}(a) The spin dephasing time as a function of the environment temperature $T$. (b) The spin dephasing time as a function of the longitudinal field gradient $b_{l}$.}
\end{figure}

The second approach is to lower the environmental temperature $T$~\cite{Culcer2009}. Lower the temperature can remarkably reduce the average occupation number $n(\omega)\approx\,k_{B}T/(\hbar\omega)$ in the low frequency noise mode. The typical temperature in experiment is about $100$ mK~\cite{yoneda2018}. The effects of lowering the temperature are shown in Fig.~\ref{fig_dephasingvstemperature}(a). The temperature dependence of the dephasing can be roughly written as $T^{*}_{2}\propto1/\sqrt{T}$. A substantial improvement in the dephasing time is achievable if the experimental temperature can be lowered to the micro-Kelvin region.

The third approach is to engineer the slanting fields~\cite{Goldman2000,Neumann2015,Yoneda2015}. As can be seen from Eqs.~(\ref{eq_edsr}) and (\ref{eq_decoherence}), the longitudinal field gradient $b_{l}$ is detrimental to both the spin manipulation and the spin dephasing. While the transverse field gradient $b_{t}$ contributes to the Rabi frequency in EDSR. Thus, it is desirable to design a proper micromagnet structure, that can give rise to both an increased transverse slanting field (shorter $T_{\rm Rabi}$) and decreased longitudinal slanting field (longer $T^{*}_{2}$). The dependence of the dephasing $T^{*}_{2}$ on the longitudinal field slope $b_{l}$ is shown in Fig.~\ref{fig_dephasingvstemperature}(b). This dependence can be roughly written as $T^{*}_{2}\propto\,1/b_{l}$.

\begin{figure}
\includegraphics[width=8.5cm]{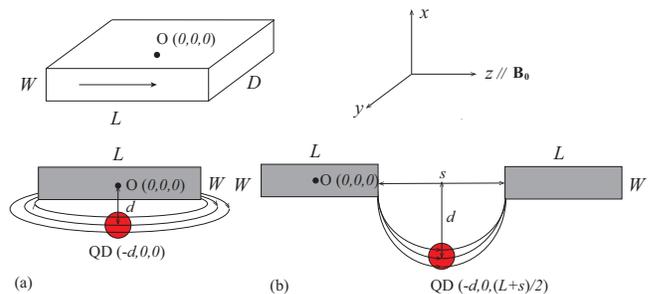}
\caption{\label{fig_micromagnet}Possible micromagnet-quantum-dot structures giving rise to reduced longitudinal field gradient ($d>W/2$).  (a) The quantum dot is placed below the micromagnetic~\cite{Goldman2000}. (b) The quantum dot is placed below two identical micromagnets~\cite{yoneda2018}. }
\end{figure}

Of great interest is designing a proper micromagnet-quantum-dot structure such that the longitudinal field gradient is reduced. Let us consider a cuboid micromagnet, the dimensions of which along $x$, $y$, and $z$ are $W$, $D$, and $L$, respectively (see Fig.~\ref{fig_micromagnet}). The external magnetic field is applied along the $z$ direction, and we assume the micromagnet is fully polarized. The origin of the coordinate system is located at the geometric center of the micromagnet. We give two possible structures with one micromagnet involved [see Fig.~\ref{fig_micromagnet}(a)] and two micromagnets involved [see Fig.~\ref{fig_micromagnet}(b)]. The $y$-dimension of the micromagnet $D$ should be large enough such that there is no $y$-component of the field ($B^{y}_{m}=0$) near the quantum dot, only $x$ and $z$-components of the field are retained ($B^{x}_{m}\neq0$ and $B^{z}_{m}\neq0$). From Eq.~(\ref{eq_slantingfield}), these ideal structures give $b_{l}=0$. 

\begin{figure}
\includegraphics{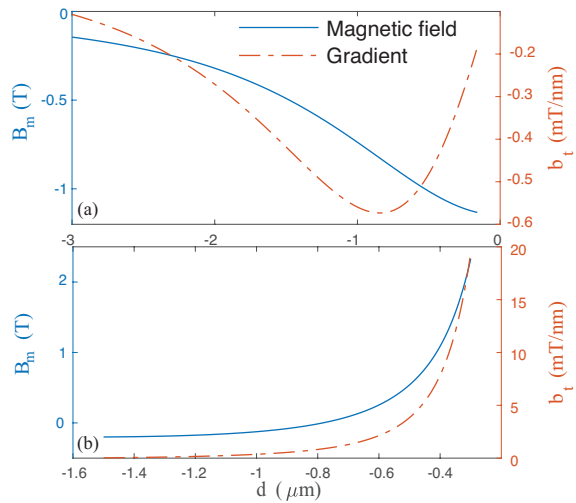}
\caption{\label{fig_micromagnet_field}The stray field and its gradient near the quantum dot. The structure parameters are $L=3$ $\mu$m, $W=0.3$ $\mu$m, $D=3.76$ $\mu$m and $s=0.2$ $\mu$m. (a) The results for single micromagnet design given in Fig.~\ref{fig_micromagnet}(a). (b) The results for two micromagnets design given in Fig.~\ref{fig_micromagnet}(b).}
\end{figure}

The Co micromagnet has a Curie temperature $T_{C}\approx1400$ K and a saturation magnetization $M_{s}=1.467\times10^{6}$ A/m~\cite{Neumann2015}. Assuming full polarization and neglecting the edge fluctuations of the micromagnet, one can obtain the field distribution using the analytical method given in Ref.~\onlinecite{Goldman2000}. Because the quantum dot is placed on the symmetrical line of the proposed micromagnet structure, from symmetry analysis, the stray field at 
$(-d,0,0)$ in Fig.~\ref{fig_micromagnet_field}(a) or $(-d,0,(L+s)/2)$ in Fig.~\ref{fig_micromagnet_field}(b) must parallel with $\hat{z}$, i.e., $\bf{B}_{m}//\hat{z}$, and its strength depends on $d$. Hence, there is a transverse field gradient $b_{t}=\partial\,B^{z}_{m}/\partial\,x=\partial\,B^{x}_{m}/\partial\,z$. While the longitudinal field gradient $b_{l}=\partial\,B^{y}_{m}/\partial\,z=\partial\,B^{z}_{m}/\partial\,y=0$ is guaranteed by the large dimension $D$ of the micromagnet. In the single micromagnet design, the maximal transverse field gradient is about $0.6$ mT/nm (see Fig.~\ref{fig_micromagnet_field}(a)). Of course, a larger field gradient is achievable by reducing $L$. In the two micromagnets design, the transverse field gradient can be as large as $10\sim20$ mT/nm (see Fig.~\ref{fig_micromagnet_field}(b)). The structure with two micromagnets more easily produces a larger transverse slanting field.

The forth promising way is to use the dynamical decoupling scheme~\cite{Uhrig2007,Lee2008,Yang2008,Cywinski2008} as has also been used in experiments. The spirit of dynamical decoupling is to frequently flip the spin using pulse sequences, such that the effective spin-noise interaction is eliminated as being of high-order small. Certainly, the performance of dynamical decoupling depends on how many pulses are applied~\cite{Medford2012}. Consider $n$ pulses applied to the qubit at a serious instant time $0<\delta_{1}t<\delta_{2}t<\ldots<\delta_{n}t<t$, i.e., at each instant time the spin qubit is flipped by the pulse, we want to determine the qubit phase coherence at the time $t$. Note that here we only consider ideal pulses, i.e., each pulse has a delta-function shape, so that the spin flip is accomplished at the instant time of the pulse applied~\cite{Uhrig2007}.

\begin{figure}
\includegraphics{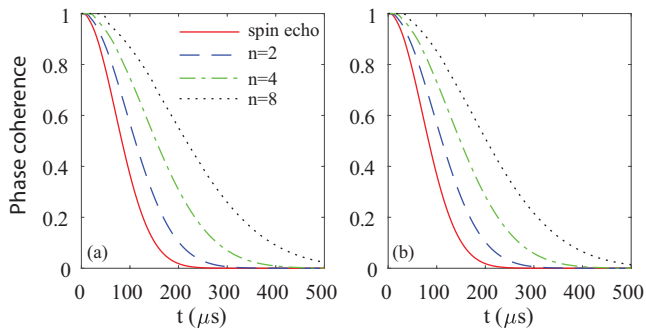}
\caption{\label{fig_dyndcoup}The phase coherence of the spin qubit under dynamical decoupling. The noise spectrum strength is chosen as $A_{r_{0}=20~{\rm nm},T=100~{\rm mK}}=35$ MHz. (a) CPMG-pulse sequences. (b) Uhrig-pulse sequences.}
\end{figure}

Under $n$-pulse sequences, the dephasing of the spin qubit due to 1/f charge noise reads as~\cite{Uhrig2007}
\begin{equation}
\Gamma^{d}_{\rm ph}(t)=\frac{y^{2}_{0}}{2r^{2}_{0}}\int^{\omega_{\rm max}}_{\omega_{\rm min}}d\omega\,S(\omega)\frac{|y_{n}(\omega\,t)|^{2}}{(\omega/2)^{2}},
\end{equation}
where
\begin{equation}
y_{n}(\omega\,t)=1+(-1)^{n+1}e^{i\omega\,t}+2\sum^{n}_{l=1}(-1)^{l}e^{i\delta_{l}\omega\,t}.
\end{equation}
For Carr-Purcell-Meiboom-Gill (CPMG) pulse sequence, the $n$ pulses are applied at the following serious instant time $\delta_{l}=(l-1/2)/n$~\cite{Carr1954,Meiboom1958}, while for Uhrig pulse sequence $\delta_{l}=\sin^{2}(\frac{\pi\,l}{2n+2})$~\cite{Uhrig2007} ($l=1,\cdots\,n$). In principle, dynamical decoupling can prolong the qubit dephasing time to any desired time scale as long as enough number of pulses are applied~\cite{Medford2012}. The practical performance of dynamical decoupling is often limited by the fact that realistic pulses are impossible in delta-function shape, i.e., the flip of the spin must cost a finite time. The phase coherence of the spin qubit under dynamical decoupling is shown in Fig.~\ref{fig_dyndcoup}.  As can be seen from the figure, the phase coherence time under spin echo is about $T^{\rm echo}_{2}\approx100$ $\mu$s. Under eight-pulse sequences, the spin dephasing time can be prolonged to $T_{2}\approx260$ $\mu$s. We also find that the CPMG-pulse sequences [see Fig.~\ref{fig_dyndcoup}(a)] perform a little better than the Uhrig-pulse sequences [see Fig.~\ref{fig_dyndcoup}(b)] in our model.

\section{Relaxation of the spin qubit}
The relaxation time $T_{1}$, i.e., the lifetime, is also an important characteristic time of the qubit~\cite{Khaetskii2001}. Even if there is no pure-dephasing for the qubit, the phase coherence time $T_{2}$ can still be limited by the qubit relaxation $T_{2}=2T_{1}$~\cite{Golovach2004}. Here we examine whether the 1/f charge noise will give rise to spin relaxation~\cite{Huang2014} in our model. The possible relaxation mechanism comes from the third term in Hamiltonian (\ref{eq_decoherence}). There is no exact method in calculating the relaxation rate, instead, the Fermi golden rule is usually used to calculate this quantity~\cite{Khaetskii2001}
\begin{equation}
\Gamma_{\rm relax}=\frac{\pi}{\hbar^{2}}\int^{\omega_{\rm max}}_{\rm \omega_{min}}\,d\omega\rho(\omega)e^{2}\Xi^{2}(z^{2}_{0}+y^{2}_{0})\delta\left(\omega-\frac{2\Delta}{\hbar}\right),
\end{equation}
where $\rho(\omega)$ is density of state of the charge noise mode. It should be noted that the qubit level spacing is about 80 GHz and the maximal charge noise frequency is about $0.5$ MHz~\cite{yoneda2018}, such that there is no charge noise frequency can match the level spacing of the spin qubit. Here arises the problem of whether the upper bound of the charge noise spectrum is indeed in the MHz range~\cite{chan2018,Kawakami2016,bylander2011,kuhlmann2013}? Our simple argument in Sec.~\ref{sec_IV} suggests $\omega_{\rm max}\ll\,k_{B}T/\hbar$ ($\sim13$ GHz for $T=100$ mK). An upper bound of 20 KHz in a SiMOS quantum dot is reported in Ref.~\onlinecite{chan2018}. Even if the qubit level spacing lies in the range of the charge noise spectrum, i.e., $\omega_{\rm min}<2\Delta/\hbar<\omega_{\rm max}$, our following calculation shows that the spin relaxation time is actually very long. By making the replacement $\sum_{k}\rightarrow\int\,d\omega_{k}\rho(\omega_{k})$ in Eq.~(\ref{eq_noisespectrum}), we have $\rho(\omega)e^{2}r^{2}_{0}\Xi^{2}\equiv\hbar^{3}A^{2}_{r_{0},T}/(k_{B}T)$. Hence, the relaxation rate can be written as
\begin{equation}
\Gamma_{\rm relax}=\frac{\pi\hbar\,A^{2}_{r_{0},T}}{k_{B}T}\times\frac{z^{2}_{0}+y^{2}_{0}}{r^{2}_{0}}.
\end{equation}
For a Si quantum dot with the parameters given in Table~\ref{tab}, we have $\Gamma_{\rm relax}=0.4519$ Hz, hence ${\rm T}_{1}=2.2$ s, indeed is a very long relaxation time. Thus, based on the above analysis, we suggest that 1/f charge noise does not limit the spin relaxation time in a Si quantum dot integrated with a slanting field.

\section{Summary}
In summary, we have studied in detail the spin manipulation and the spin dephasing in a Si quantum dot integrated with a slanting magnetic field. The longitudinal slanting field not only gives rise to a modulated Rabi oscillation in the spin manipulation, but also mediates a longitudinal spin-charge interaction which leads to spin dephasing. Several practical strategies are also proposed to alleviate the spin dephasing. Also, 1/f charge noise does not limit the spin relaxation time due to the mismatching between the qubit level spacing and the charge noise frequency. Our study can help clarify the spin dephasing mechanism in Si quantum dot.

\section*{Acknowledgements}
This work is supported by the National Natural Science Foundation of China Grant No.~11404020, the Postdoctoral Science Foundation of China Grant No.~2014M560039, and Doctoral Fund of Yanshan University Grant No. BL18043.

\appendix
\section{\label{appendix_A}The basis states of the spin qubit Hilbert space}
In this appendix, the quantum states which span the qubit Hilbert space will be studied using perturbation theory. The quantum dot Hamiltonian can be divided into two parts:
\begin{eqnarray}
H&=&H_{0}+H',\nonumber\\
H_{0}&=&\frac{p^{2}_{y}+p^{2}_{z}}{2m}+\frac{m\omega^{2}_{0}}{2}(y^{2}+z^{2}+2y_{0}y\sigma^{z})+\Delta\sigma^{z},\nonumber\\
H'&=&m\omega^{2}_{0}z\sqrt{z^{2}_{0}+y^{2}_{0}}(\sigma^{x}\cos\theta_{0}+\sigma^{y}\sin\theta_{0}),
\end{eqnarray}
where $H_{0}$ is quasi-diagonalized due to the factor that operator $\sigma^{z}$ is conserved, and $H'$ will be regarded as a perturbation in our following calculation. The lowest two energy levels in the quantum dot are used to encode a qubit, such that we only need to calculate the eigenvalues and the corresponding eigenfunctions for the lowest two energy levels.  The zeroth-order eigenvalues read
\begin{eqnarray}
E^{(0)}_{n_{y},n_{z},\uparrow}&=&(n_{y}+n_{z}+1)\hbar\omega_{0}+\Delta-\frac{m\omega^{2}_{0}}{2}y^{2}_{0},\nonumber\\
E^{(0)}_{n_{y},n_{z},\downarrow}&=&(n_{y}+n_{z}+1)\hbar\omega_{0}-\Delta-\frac{m\omega^{2}_{0}}{2}y^{2}_{0}.~~~~~
\end{eqnarray}
The corresponding zeroth-order eigenfunctions read
\begin{eqnarray}
|\Psi^{(0)}_{n_{y},n_{z},\uparrow}\rangle&=&D(-y_{0})|n_{y},n_{z},\uparrow\rangle,\nonumber\\
|\Psi^{(0)}_{n_{y},n_{z},\downarrow}\rangle&=&D(y_{0})|n_{y},n_{z},\downarrow\rangle,
\end{eqnarray}
where $D(y_{0})={\rm exp}(-iy_{0}p_{y}/\hbar)$ is the displacement operator in the $y$ dimension, and $|n_{y},n_{z},\sigma\rangle$ is the eigenfunction of the bare harmonic oscillator
\begin{eqnarray}
&&\left(\frac{p^{2}_{y}+p^{2}_{z}}{2m}+\frac{m\omega^{2}_{0}}{2}(y^{2}+z^{2})+\Delta\sigma^{z}\right)|n_{y},n_{z},\sigma\rangle=\nonumber\\
&&~~~~~~~~~\big[(n_{y}+n_{z}+1)\hbar\omega_{0}+\sigma\Delta\big]|n_{y},n_{z},\sigma\rangle.
\end{eqnarray}
By using the first-order non-degenerate perturbation formula~\cite{Landau1965}, we obtain the eigenfunctions
\begin{eqnarray}
|\Psi_{0,0,\uparrow}\rangle&=&|\Psi^{(0)}_{0,0,\uparrow}\rangle-m\omega^{2}_{0}\sqrt{z^{2}_{0}+y^{2}_{0}}e^{i\theta_{0}}e^{-y^{2}_{0}/r^{2}_{0}}\times\nonumber\\
&&\sum^{\infty}_{n_{y}=0}\frac{(-1)^{n_{y}}2^{\frac{n_{y}-1}{2}}y^{n_{y}}_{0}}{\sqrt{n_{y}!}r^{n_{y}-1}_{0}\big[(n_{y}+1)\hbar\omega_{0}-2\Delta\big]}|\Psi^{(0)}_{n_{y},1,\downarrow}\rangle,\nonumber\\
|\Psi_{0,0,\downarrow}\rangle&=&|\Psi^{(0)}_{0,0,\downarrow}\rangle-m\omega^{2}_{0}\sqrt{z^{2}_{0}+y^{2}_{0}}e^{-i\theta_{0}}e^{-y^{2}_{0}/r^{2}_{0}}\times\nonumber\\
&&\sum^{\infty}_{n_{y}=0}\frac{2^{\frac{n_{y}-1}{2}}y^{n_{y}}_{0}}{\sqrt{n_{y}!}r^{n_{y}-1}_{0}\big[(n_{y}+1)\hbar\omega_{0}+2\Delta\big]}|\Psi^{(0)}_{n_{y},1,\uparrow}\rangle.\nonumber\\
\end{eqnarray}
The corresponding first-order perturbation eigen-energies read
\begin{equation}
E_{0,0,\uparrow}=E^{(0)}_{0,0,\uparrow},~~~E_{0,0,\downarrow}=E^{(0)}_{0,0,\downarrow}.
\end{equation}
Thus, the first-order perturbation gives no corrections to the energies.

When an in-plane oscillating electric-field is applied to the quantum dot, there is an electric-dipole interaction $e\textbf{E}\cdot\textbf{r}\cos(\omega\,t)$ between the electron and the driving field. In the qubit Hilbert space, we can calculate the matrix elements for the coordinate operator $y$:
\begin{eqnarray}
&&\langle\Psi_{0,0,\uparrow}|y|\Psi_{0,0,\uparrow}\rangle=-y_{0},~\langle\Psi_{0,0,\downarrow}|y|\Psi_{0,0,\downarrow}\rangle=y_{0},\nonumber\\
&&~~~~~~~~~~~~~~~~~~~\langle\Psi_{0,0,\uparrow}|y|\Psi_{0,0,\downarrow}\rangle=0.
\end{eqnarray}
Hence the operator $y$ can be written as
\begin{equation}
y=-y_{0}\tau^{z}.
\end{equation}
We also can calculate the matrix elements for the coordinate operator $z$:
\begin{eqnarray}
&&\langle\Psi_{0,0,\uparrow}|z|\Psi_{0,0,\uparrow}\rangle=0,~\langle\Psi_{0,0,\downarrow}|z|\Psi_{0,0,\downarrow}\rangle=0,\nonumber\\
&&\langle\Psi_{0,0,\uparrow}|z|\Psi_{0,0,\downarrow}\rangle=-\sqrt{z^{2}_{0}+y^{2}_{0}}e^{-i\theta_{0}}e^{-y^{2}_{0}/r^{2}_{0}}\frac{\hbar^{2}\omega^{2}_{0}}{\hbar^{2}\omega^{2}_{0}-4\Delta^{2}}.
\end{eqnarray}
Under the realistic parameter condition (see Table.~\ref{tab}), $y_{0}\ll\,r_{0}$ and $2\Delta\ll\hbar\omega_{0}$, such that the operator $z$ can be written as
\begin{equation}
z=-\sqrt{z^{2}_{0}+y^{2}_{0}}(\tau^{x}\cos\theta_{0}+\tau^{y}\sin\theta_{0}).
\end{equation}
Therefore, under electric-field driving, the quantum dot Hamiltonian can be written as the form given by Eq.~(\ref{eq_edsr}). Note that $\langle\Psi_{0,0,\uparrow}|z|\Psi_{0,0,\downarrow}\rangle$, which contributes to the Rabi frequency in the EDSR, is almost independent of the Zeeman field $B_{0}$~\cite{Hu2012}.

\bibliographystyle{iopart-num}
\bibliography{ChargeNoiseRef}
\end{document}